\documentclass[twocolumn,prl,showpacs,amsfonts,amsmath,assymb,eufrak]{revtex4}
\def\eq#1\en{\begin{equation}#1\end{equation}}  
\def\eqa#1\ena{\begin{align}#1\end{align}}
\def\eqg#1\eng{\begin{gather}#1\end{gather}}
\newcommand{\lb}[1]{\label{e:#1}}
\newcommand{\rlb}[1]{\eqref{e:#1}} 
\newcommand{\nl}{\notag\\}

\newcommand{\sbkt}[1]{\langle#1\rangle}

\newcommand{\sumtwo}[2]%
{\mathop{\sum_{#1}}_{#2}}
\newcommand{\sumthree}[3]%
{\mathop{\mathop{\sum_{#1}}_{#2}}_{#3}}
\newcommand{\sumfour}[4]%
{\mathop{\mathop{\mathop{\sum_{#1}}_{#2}}_{#3}}_{#4}} 
\newcommand{\prodtwo}[2]%
{\mathop{\prod_{#1}}_{#2}}
\newcommand{\mintwo}[2]%
{\mathop{\min_{#1}}_{#2}}
\newcommand{\maxtwo}[2]%
{\mathop{\max_{#1}}_{#2}}
\newcommand{\maxthree}[3]%
{\mathop{\mathop{\max_{#1}}_{#2}}_{#3}}
\newcommand{\limtwo}[2]%
{\mathop{\lim_{#1}}_{#2}}
\newcommand{\suptwo}[2]%
{\mathop{\sup_{#1}}_{#2}}
\newcommand{\supthree}[3]%
{\mathop{\mathop{\sup_{#1}}_{#2}}_{#3}}
\newcommand{\supfour}[4]%
{\mathop{\mathop{\mathop{\sup_{#1}}_{#2}}_{#3}}_{#4}} 
\newcommand{\inftwo}[2]%
{\mathop{\inf_{#1}}_{#2}}
\newcommand{\infthree}[3]%
{\mathop{\mathop{\inf_{#1}}_{#2}}_{#3}}
\newcommand{\inffour}[4]%
{\mathop{\mathop{\mathop{\inf_{#1}}_{#2}}_{#3}}_{#4}} 
\newcommand\calG{{\cal G}}

\newcommand{\ep}{\epsilon}
\newcommand{\up}{\uparrow}

\newcommand{\phii}{\varphi^\mathrm{init}}
\newcommand{\phif}{\varphi^\mathrm{fin}_\tau}
\newcommand{\hH}{\hat{H}}
\newcommand{\hU}{\hat{U}}
\newcommand{\hP}{\hat{P}_0}
\newcommand{\UPU}{\hat{U}^\dagger\hat{P}_0\hat{U}}
\newcommand{\Pt}{P_\tau[\hH\le E_0-V\delta]}
\newcommand{\para}[1]{{\em #1}\/.---}

\begin{document}
\title{The second law of thermodynamics for pure quantum states}

\author{Sheldon Goldstein${}^1$, Takashi Hara${}^2$, and Hal Tasaki${}^3$}
\affiliation{
${}^1$%
Departments of Mathematics and Physics, Rutgers University, 110 Frelinghuysen Road, Piscataway, NJ 08854-8019, USA
\\
${}^2$%
Faculty of Mathematics,
Kyushu Univerisity,
Moto-oka, Nishi-ku,
Fukuoka 819-0395,
Japan
\\
${}^3$%
Department of Physics, Gakushuin University, 
Mejiro, Toshima-ku, Tokyo 171-8588, Japan}

\date{\today}

%%%%%%%%%%%%%%%%%
\begin{abstract}
A version of the second law of thermodynamics states that one cannot lower the energy of an isolated system by a cyclic operation.
We prove this law without introducing statistical ensembles and by resorting only to quantum mechanics.
We choose the initial state as a pure quantum state whose energy is almost $E_0$ but not too sharply concentrated at energy eigenvalues.
Then after an arbitrary unitary time evolution which follows a typical ``waiting time'', the probability of observing the energy lower than $E_0$ is proved to be negligibly small.
\end{abstract}

% 05.30.-d 	Quantum statistical mechanics 
%05.70.-a 	Thermodynamics
% 03.65.Yz 	Decoherence; open systems; quantum statistical methods

\maketitle
%%%%%%%%%%%%%%%%%%%%%%%%%%%%%%%%%%%%%
%%%%%%%%%%%%%%%%%%%%%%%%%%%%%%%%%%%%%

The recent renewed interest in the foundation of quantum statistical mechanics and in the dynamics of isolated quantum systems has led to a revival of the old idea  that a single pure quantum state can describe a thermal equilibrium state \cite{vonNeumann}.
Although our understanding of quantum mechanics of fully interacting many-body  quantum systems is still primitive, it has become clear that pure quantum states in suitable classes generically describe equilibrium states \cite{vonNeumann,Hal1998,PopescuShortWinter,GLTZ06, Reimann,LindenPopescuShortWinter,GLMTZ09b,Hal2010,Reimann2,SugiuraShimizu}.

In the present work, we shall go further and prove that, in an abstract (but physically natural) setting, a pure quantum state and quantum dynamics alone lead to thermodynamics, in particular, the second law \cite{Hal2000,Ikeda}.
Although various methods to derive the second law from quantum dynamics are known \cite{Lindblad,PuszWoronowicz,Lenard,Kurchan,Hal2000b,Sagawa}, they all rely essentially on the fact that the initial state is chosen as a mixed state that corresponds to a statistical mechanical ensemble.
In the present derivation, we deal only with pure states but make a full use of the fact that the system is macroscopic.

Among several expressions of the second law, we focus on the most fundamental one, which asserts that {\em one cannot lower the energy of an isolated system by a cyclic operation}\/.
The statement is called passivity in the mathematical literature \cite{PuszWoronowicz}, and sometimes called Planck's principle in the context of thermodynamics \cite{LiebYngvason}.
Consider a thermodynamic system which is initially in an equilibrium state with energy $U_\mathrm{init}$.
We assume that the system is isolated and does not exchange heat with the environment.
Then an external agent performs a {\em mechanical}\/ operation to the system by changing controllable parameters (such as the position of a piston or the direction and the magnitude of an external field).
We also assume that the operation is {\em cyclic}\/ in the sense that all the parameters return to their initial values at the end of the operation.
Then the second law of thermodynamics states that $U_\mathrm{init}\le U_\mathrm{fin}$, where $U_\mathrm{fin}$ is the energy of the final state \cite{entropy}.
Other forms of the second law may be derived from this postulate by using appropriate auxiliary assumptions.

Here we shall model the above situation faithfully in terms of quantum mechanics.
The initial state is chosen to be a pure state whose energy is almost $E_0$ but not sharply peaked at any energy eigenvalues.
The operation is modeled as a fixed unitary time evolution $\hU$ which follows a ``waiting time'' $\tau$.
Then we prove that, for a typical choice of $\tau$, the probability of observing the energy lower than $E_0$ in the final state is essentially vanishing.

We hope that this sharp and rigorous result sheds light on the foundation of thermodynamics and statistical mechanics, and also leads to a deeper understanding of the relation between microscopic quantum theory and macroscopic physics.

\para{Setup and the main result}%
We consider an abstract model for a macroscopic quantum system confined in a finite volume.  
The volume $V$ is here treated as a fixed parameter that characterizes the system.
A typical example is a particle system with a constant density.

Let $\hH$ be the Hamiltonian, whose $V$ dependence is omitted.
For $j=1,2,\ldots$, we denote by $E_j$ and $\psi_j$ the eigenvalue and  the normalized eigenstate, respectively, of $\hH$, i.e., $\hH\psi_j=E_j\psi_j$.
We assume that the energy eigenvalues are nondegenerate, and ordered so that $E_j<E_{j+1}$.
The absence of degeneracy is the only assumption for $\hH$.

As usual we denote by $\Omega_V(E)$ the number of energy eigenstates such that $E_j\le E$.
We assume that there is a  strictly increasing function $\sigma(\ep)$ which is independent of $V$, and one has
\eq
\Omega_V(E)=\exp[V\sigma(E/V)+o(V)],
\lb{OV}
\en
which is indeed the property usually found in macroscopic quantum systems \cite{Ruelle}.

We assume that an outside agent performs an operation to the system by changing the Hamiltonian in an arbitrary manner.
We require the operation to be cyclic in the sense that the initial and the final Hamiltonian are the same $\hH$ \cite{cyclic}.
We shall fix such an operation, and denote by $\hU$ the unitary time evolution for the whole operation.
Theoretically speaking, $\hU$ is treated here as an arbitrary fixed unitary operator.

We wish to investigate whether the time evolution defined by $\hU$ leads to the second law of thermodynamics.
Obviously this is impossible in general.
For an arbitrary initial state $\phii$ with an arbitrarily high energy, one can always (at lest formally) find a unitary operator $\hU$ such that the final state $\varphi^\mathrm{fin}=\hU\phii$ is, say, the ground state of $\hH$.
Then the energy of the final state is of course much lower than that of the initial state; the second law is  drastically violated.

In order to avoid the use of such a ``custom-made'' time-evolution, we introduce a little twist.
For an initial state $\phii$, we consider the final state defined as $\phif:=\hU e^{-i\hH\tau}\phii$.
Here $\tau>0$ is the ``waiting time'', during which the system simply evolves according to the constant Hamiltonian $\hH$.
By choosing the waiting time $\tau$ randomly (for a fixed $\hU$), we can inhibit the time-evolution from making full use of the detailed properties of the initial state.
From a physical point of view, the introduction of the random waiting time may be regarded as a natural expression of the fact that we never have a perfect control on exactly when an operation starts in the laboratory.

We shall choose the initial state $\phii$ from the ``energy shell'' including an arbitrary energy $E_0$.
More precisely $\phii=\sum_j\alpha_j\psi_j$ where $\alpha_j$ is nonvanishing only when $E_j/V\in[\ep_0,\ep_0+\delta]$, where $\ep_0:=E_0/V$ and $\delta$ is a small constant independent of $V$.
The most crucial assumption \cite{assumption} is that the coefficients satisfy
\eq
|\alpha_j|^2\le\frac{1}{\Omega_V(E_0)},
\lb{aj}
\en 
which roughly says that the state $\phii$ does not have too sharp peaks at  precise energy eigenstates.
Note that the total number of the energy eigenstates (i.e., the dimension) in the present energy shell is $\Omega_V(E_0+V\delta)-\Omega_V(E_0)\simeq\Omega_V(E_0+V\delta)$, which is exponentially larger than $\Omega_V(E_0)$ for large $V$.
This means that a typical state in the energy shell indeed satisfies the condition \rlb{aj}.

Now suppose that one makes a projective measurement of the energy $\hH$ in the final state $\phif$.
We denote by $\Pt$ the probability that the outcome is less than or equal to $E_0-V\delta$.
Then our main result is as follows.

\para{Theorem}%
For an arbitrary $\phii$ and the operation (or evolution) $\hU$, there are positive constants $T$ and $a\simeq\{\sigma(\ep_0)-\sigma(\ep_0-\delta)\}/2$, and we have the following.
There exists a ``good'' subset $\calG\subset[0,T]$ whose total length $|\calG|$ satisfies 
$|\calG|/T\ge1-e^{-a V}$. 
For any $\tau\in\calG$, one has $\Pt\le e^{-a V}$.

Note that $e^{-aV}$ is negligibly small for large $V$.
Thus, for a macroscopic system, the theorem says that the probability of observing the energy lower than $E_0-V\delta$ in the final state is essentially vanishing for a sufficiently long and typical waiting time $\tau\in[0,T]$.
Since the energy of the initial state is in between $E_0$ and $E_0+V\delta$, this shows that the observed energy in the final state can be lower than the initial energy only by $O(V\delta)$.
Noting that $V\delta$ can be chosen negligibly small in a macroscopic system, we see that the theorem  establishes the the second law in the sense that one cannot lower the energy of an isolated macroscopic system by a cyclic process.

The second law stated in the theorem is stronger than the standard statement that refers to the energy expectation value in the final state \cite{Hal2000,Lenard,Sagawa}.
Note that, when the final state happens to be a superposition of states with macroscopically distinct energies (i.e., a Schr\"{o}dinger's cat), the energy expectation value has little physical meaning.
Clearly our theorem implies the lower bound for the expectation value, i.e.,
\eq
\sbkt{\phif,\hH\phif}\ge (1-e^{-aV})(E_0-V\delta)\simeq E_0.
\en

\para{Proof}%
From the definitions the final state is
\eq
\phif=\sum_j\alpha_j e^{-iE_j\tau}\hU\psi_j.
\en
Let $\hP$ be the orthogonal projection onto the space with $\hH\le E_0-V\delta$.
(In Dirac notation $\hP=\sum_{k\,(\mathrm{s.t.}\,E_k\le E_0-V\delta)}|\psi_k\rangle\langle\psi_k|$.)
Then we have
\eqa
&\Pt=\sbkt{\phif,\hP\phif}
\nl
&\hspace{20pt}=\sum_{j,k}\alpha_j\alpha^*_ke^{-i(E_j-E_k)\tau}
\sbkt{\psi_k,\UPU\psi_j}.
\lb{Pt}
\ena
Let us abbreviate $\Pt$ as $P_\tau$ in what follows.
For any function $f_\tau$ of the waiting time $\tau>0$, define its time average as $[f_\tau]_T:=T^{-1}\int_0^Td\tau f_\tau$, where $T>0$.
The $T\up\infty$ limit is written as $[f_\tau]_\infty$.
Since energy eigenvalues are nondegenerate, one readily finds
\eqa
[P_\tau]_\infty&=\sum_j|\alpha_j|^2\sbkt{\psi_j,\UPU\psi_j}
\nl
&\le\max_j|\alpha_j|^2\sum_{j=1}^\infty\sbkt{\psi_j,\UPU\psi_j}
\nl
&=\max_j|\alpha_j|^2\,\mathrm{Tr}[\UPU]
%\nl&
=\frac{\Omega_V(E_0-V\delta)}{\Omega_V(E_0)}
\nl&
=\exp[V\{\sigma(\ep_0-\delta)-\sigma(\ep_0)\}+o(V)]
\nl&
\le e^{-2aV}/2,
\lb{Pinf}
%\lb{Ptinf}
\ena
with $a\simeq\{\sigma(\ep_0)-\sigma(\ep_0-\delta)\}/2>0$, where we used the bound \rlb{aj}.

The bound \rlb{Pinf} implies $[P_\tau]_T\le e^{-2aV}$ for sufficiently large $T$.
We fix such $T$, and define the ``good'' subset as
\eq
\calG=\bigl\{\tau\in[0,T]\,\bigr|\,P_\tau\le e^{-aV}\,\bigr\}.
\lb{G}
\en
Let $I[\text{true}]=1$ and $I[\text{false}]=0$.
We have
\eqa
e^{-2aV}&\ge[P_\tau]_T\ge\bigl[P_\tau I[P_\tau>e^{-aV}]\bigr]_T
\nl&
\ge e^{-aV}\bigl[I[P_\tau>e^{-aV}]\bigr]_T
=e^{-aV}\Bigl(1-\frac{|\calG|}{T}\Bigr),
\ena
which implies the desired bound for $|\calG|$.

\para{Other formulations}%
We have presented the second law as a statement for a typical waiting time, but this setting is not mandatory.
One can also talk about a typical operation as the following.

Again fix an arbitrary unitary operator $\hU$.
Take $\theta_j\in[0,2\pi)$ for each $j=1,2,\ldots$, and define ${\hU}_{\boldsymbol{\theta}}$ by $\sbkt{\psi_k,{\hU}_{\boldsymbol{\theta}}\psi_j}=e^{i\theta_j}\sbkt{\psi_k,\hU\psi_j}$ for any $k$ and $j$.
Then we consider the evolution by the unitary operator ${\hU}_{\boldsymbol{\theta}}$ without a waiting time.
We here assume that the parameters $\theta_j$ are chosen randomly with a uniform probability.
Then by exactly the same proof, we can show the statements of the theorem for a typical choice of $\theta_j$ (with $j=1,2,\ldots$).

This formulation may be regarded as a mathematical expression of the fact that we never have a perfect control on the operation that determines the unitary evolution.
The question remains, however, as to whether the uniform measure on $[0,2\pi)$ is ``realistic''.
We stress that the uniform measure of the waiting time is the unique choice because of the translation invariance of time.

Note that using ${\hU}_{\boldsymbol{\theta}}$ is equivalent to considering the initial state $\phii=\sum_j e^{i\theta_j}|\alpha_j|\psi_j$ with random $\theta_j$ for a fixed $\hU$.
One can go further in this direction to show the second law as a statement for a typical state in the energy shell.
First fix the unitary operator $\hU$, then choose the initial state $\phii$ uniformly from the energy shell, and finally let $\varphi^\mathrm{fin}=\hU\phii$.
Then the second law holds with a probability close to 1.
Note that we don't have to assume the condition \rlb{aj} in this formulation.

In the present work, we have imposed the essential condition \rlb{aj} to the initial state.
Although the condition is satisfied by a typical state in the energy shell, one can still ask if a statement for {\em any}\/ initial state is possible.
We recall that, in the problem of ``approach to equilibrium'', it was shown (in some settings) that an arbitrary initial state in an energy shell approaches the corresponding thermal equilibrium state \cite{vonNeumann,GLMTZ09b,Hal2010}.

A simple strategy to state such a claim is to assume that the unitary operator $\hU$ satisfies the bound
\eq
|\sbkt{\psi_k,\hU\psi_j}|^2\le\frac{1}{\Omega_V(E_j)},
\lb{Ukj}
\en
which roughly means that $\hU$ sufficiently ``shuffles'' the energy eigenstates around $E_j$.
Again by considering the time evolution $\hU e^{-i\tau\hH}$ with a waiting time, and noting that $[P_\tau]_\infty=\sum_{j,k\,({\rm s.t.} E_k\le E_0-V\delta)}|\alpha_j|^2|\sbkt{\psi_k,\hU\psi_j}|^2$, we get the same second law for any initial state $\phii$ from the energy shell.
We still do not know how restrictive (or realistic) the condition \rlb{Ukj} is.

Another somewhat artificial strategy is to combine two unitary operators.
Let  $\hU_\mathrm{s}$ be the ``shuffling" unitary operator with the properties that $\sbkt{\psi_k,\hU_\mathrm{s}\psi_j}$ is nonvanishing only when $E_k\sim E_j$, and $|\sbkt{\psi_k,\hU_\mathrm{s}\psi_j}|^2\le 1/\Omega_V(E_j)$.
Such an operator which only ``shuffles'' nearby energy eigenstates may be constructed for a general Hamiltonian.
We first apply $\hU_\mathrm{s}$ and then an arbitrary $\hU$ which corresponds to the thermodynamic operation.
We also need two independent waiting times before and after $\hU_\mathrm{s}$.
Thus the whole unitary evolution is $\hU e^{-iH\tau}\hU_\mathrm{s}e^{-iH\tau'}$.
Then for typical choices of $\tau$ and $\tau'$, we can show the statement corresponding to the second law for an arbitrary initial state $\phii$ from the energy shell.
The trick is that $\hU_\mathrm{s}$ (along with the typicality in $\tau'$) modifies an arbitrary state into one satisfying the condition \rlb{aj}.

Although the above construction is artificial, it could be that a realistic time evolution in a macroscopic system has some features in common with our shuffling operator $\hU_\mathrm{s}$.
To develop a more realistic theory following this philosophy is an interesting challenge.

Let us note in passing that it is trivial to prove the same second law when the initial state is a mixed state.
More precisely if  $\hat{\rho}_\mathrm{init}=\sum_j|\alpha_j|^2|\psi_j\rangle\langle\psi_j|$ with the condition \rlb{aj}, then $P[\hH\le E_0-V\delta]:={\rm Tr}[\hP\hU\hat{\rho}_\mathrm{init}\hU^\dagger]\le e^{-2a V}/2$, since ${\rm Tr}[\hP\hU\hat{\rho}_\mathrm{init}\hU^\dagger]=[P_\tau]_\infty$.

\para{Application to the problem of ``approach to equilibrium''}%
Note that our proof only makes use of the fact that $P_0$ is an orthogonal projection onto a space whose dimension is much smaller than $\Omega_V(E_0)$.
This means that we can apply exactly the same technique to the problem of approach to equilibrium by a time evolution with fixed $\hH$ \cite{vonNeumann,Hal1998,Reimann,LindenPopescuShortWinter,
GLMTZ09b,Hal2010,Reimann2}.
Then $\tau$ should be interpreted as the time required for the relaxation 

In this problem we set $\hU=1$ and let $\hP=1-\hat{P}_\mathrm{eq}$, where $\hat{P}_\mathrm{eq}$ is the orthogonal projection onto the properly defined ``equilibrium subspace'' (see \cite{GLMTZ09b}) within the energy shell (spanned by $\psi_j$ such that $E_j/V\in[\epsilon_0,\epsilon_0+\delta]$).
The initial state is assumed to satisfy the same condition as in the main theorem including the crucial \rlb{aj}.

Then, exactly as in \rlb{Pinf}, we find
\eq
[\sbkt{\phif,(1-\hat{P}_\mathrm{eq})\phif}]_\infty
\le\frac{d_\mathrm{neq}}{\Omega_V(E_0)},
\en
where $d_\mathrm{neq}$ is the codimension of the equilibrium subspace (i.e., the dimension of the ``nonequilibrium subspace'').
When one has $d_\mathrm{neq}\ll\Omega_V(E_0)$, this leads to a theorem that  the state $\phif=e^{-iH\tau}\phii$ is ``very close'' to the equilibrium subspace for a sufficiently long and typical $\tau$ as in \cite{GLMTZ09b}.
Note that this theorem does not require any assumptions on the energy eigenstates $\psi_j$.

\para{Discussions}%
We have proved a theorem in quantum mechanics that essentially establishes the second law of thermodynamics in a form directly relevant to (thermodynamic) experiments.
We hope that this rigorous and sharp result sheds light on the connection between microscopic and macroscopic physics.

Although it is sometimes argued that the macroscopic irreversibility expressed in the second law is in conflict with the unitary time evolution in quantum mechanics, our theorem clearly shows that this is not the case.
Let us discuss essential ingredients in our theory.

From a mathematical point of view, a crucial observation is that the long-time average of the relevant quantity is expressed as in the first line of Eq.~\rlb{Pinf}, which can be interpreted as an expression in {\em classical}\/ stochastic process (which starts from a state $j$ and ends in $k$ such that $E_k\le E_0$).
Then the second law is almost trivial under the assumption \rlb{aj}.

It is clear that the bound \rlb{aj} is the most important (and essentially the unique) assumption in the present theory.
Let us again stress that this is quite a natural requirement since a state preparation system in general does not have precise information about the energy eigenstates of a macroscopic system, or to put it the other way round, it is an almost impossible task for an experimentalist to prepare an initial state that violates the bound \rlb{aj}.
Nevertheless we still do not fully understand whether the condition \rlb{aj} (or its equivalent like \rlb{Ukj} or the one for the shuffling operator $\hU_\mathrm{s}$ above) is really indispensable for the derivation of the second law.
Clearly this is the most delicate assumption in the present work; one might even argue that it is in a sense parallel to introducing a probability distribution ``by hand''.
All that we can say for the moment is that such a condition is necessary if one follows the philosophy as in the present paper.

The introduction of the random ``waiting time'' $\tau$ is an essential theoretical trick to exclude very special unitary time evolution which accidentally lowers the energy of the system.
As we have already stressed this is quite a natural notion since one never has a perfect control on the operation to a physical system.
We have assumed that $\tau$ is drawn from the uniform distribution on the interval $[0,T]$, but this seems to be the unique natural choice of the measure.

We must remark however that our general theory does not provide us with any quantitative estimate of $T$, the upper limit of the waiting time.
Physically speaking, the time necessary to exclude very special unitary evolution is expected to be quite short, but should depend crucially on the nature of the system and the initial state.
It is likely that we have to develop much more concrete and specific theories to deal with the problem of time scale.

The typicality of equilibrium states \cite{PopescuShortWinter,GLTZ06} suggests that almost all initial states $\phii$ that satisfy our assumptions correspond to the thermal equilibrium state with energy $E_0$.
But there are some $\phii$ (satisfying our assumptions) describing macroscopic states which are very far from equilibrium.
It is interesting that our theorem is equally valid for such nonequilibrium initial states.

When the initial state is out of equilibrium, one may interpret the waiting time $\tau$ as the relaxation time necessary for the state to reach the equilibrium, but we are not sure whether this interpretation is mandatory.
It is also likely that  the most important role of the waiting time is to inhibit the operation from (accidentally) making use of special features of the initial state.
This suggests an interesting (and probably novel) view that the most essential requirement for the second law is the lack of information about the initial state (rather than the equilibrium nature of the initial state), and the second law holds for a much larger class of initial states than equilibrium states.

It is a pleasure to thank
Tatsuhiko N. Ikeda,
Takahiro Sagawa,
Keiji Saito,
Shin-ichi Sasa,
and
Akira Shimizu
for valuable discussions.

%%%%%%%%%%%%%%%%%%%%%%%%%%%%%%%%%%%%%%%%%%%
%%%%%%%%%%%%%%%%%%%%%%%%%%%%%%%%%%%%%%%%%%%

%%%%%%%%%%%%%%%%%%%%%%%%%%%%%%%%%%%%%%
%%%%%%%%%%%%%%%%%%%%%%%%%%%%%%%%%%%%%%
%%%%%%%%%%%%%%%%%%%%%%%%%%%%%%%%%%%%%%

\begin{thebibliography}{10}
\bibitem{entropy}
An essential advantage of this formulation of the second law is that it does not refer to the entropy.
Recall that, in thermodynamics, the entropy is not a directly observable quantity, while the change in the energy is.
By focusing only on the energy, we can develop a microscopic theory that is free from arbitrariness of the (microscopic) definition of entropy.
\bibitem{cyclic}
The cyclicity is assumed only for simplicity.
The initial and the final Hamiltonians may be distinct (as in \cite{Hal2000}), but then the setting becomes complicated.
\bibitem{assumption}
The bound \rlb{aj} is the only assumption we need for the initial state.
The assumption that $\phii$ is taken from the energy shell is made only to clarify the physical situation, and is not necessary.
\end{thebibliography}
\end{document}